\def\slash#1{\setbox0=\hbox{$#1$}#1\hskip-\wd0\dimen0=5pt\advance
\dimen0 by-\ht0\advance\dimen0 by\dp0\lower0.5\dimen0\hbox
to\wd0{\hss\sl/\/\hss}}
\def\Black{}
\def\Blue{}
\def\Brown{}

\documentstyle[aps,epsf,prl]{revtex}
\newcommand{\be}{\begin{equation}}
\newcommand{\ee}{\end{equation}}
\newcommand{\dd}{\displaystyle}

\begin{document}

\begin{titlepage}
\null
\vspace{2cm}
\begin{center}
\Large\bf 
\Brown
Semileptonic $B \to \rho$ and $B \to a_1$ transitions in a 
quark-meson model
\Black
\end{center}
\vspace{1.5cm}

\begin{center}
\begin{large}
A. Deandrea\\
\end{large}
\vspace{0.5cm}
Institut f\"ur Theoretische Physik, Universit\"at Heidelberg,\\
Philosophenweg 19, D-69120 Heidelberg, Deutschland\\
\vspace{0.7cm}
\begin{large}
R. Gatto\\
\end{large}
\vspace{0.5cm}
D\'epartement de Physique Th\'eorique, Universit\'e de Gen\`eve,\\
24 quai E.-Ansermet, CH-1211 Gen\`eve 4, Suisse\\
\vspace{0.7cm}
\begin{large}
G. Nardulli and A. D. Polosa\\
\end{large}
\vspace{0.5cm}
Dipartimento di Fisica, Universit\`a di Bari and INFN Bari,\\
via Amendola 173, I-70126 Bari, Italia
\end{center}

\vspace{1.3cm}

\begin{center}
\begin{large}
\Brown
{\bf Abstract}\\[0.5cm]\Black
\end{large}
\parbox{14cm}{We evaluate the form factors governing the exclusive decays
$B\to\rho\ell\nu$, $B\to a_1 \ell\nu$, by using an effective quark-meson
lagrangian. The model  is based on meson-quark interactions, and the 
computation of the mesonic transition amplitudes is performed by
considering  diagrams with heavy mesons attached to loops containing heavy 
and light constituent quarks. This approach
was successfully employed to compute the 
Isgur-Wise form factors and other hadronic observables for negative and 
positive parity heavy mesons and is presently used for exclusive 
heavy-to-light weak transitions. We also evaluate a few strong coupling 
constants appearing in chiral effective lagrangians for heavy and light 
mesons.}
\end{center}

\vspace{1.5cm}
\noindent
\Blue
\Black
\vfil
\noindent
\Brown
BARI-TH/98-314\\
HD-TVP-98-8\\
UGVA-DPT 1998/11-1020\\
November 1998
\Black
\end{titlepage}

\setcounter{page}{1}


\preprint{BARI-TH/98-314\\
HD-TVP-98-8\\
UGVA-DPT 1998/11-1020}
\title{Semileptonic $B \to \rho$ and $B \to a_1$ transitions in a 
quark-meson model}
\author{A. Deandrea}
\address{Institut f\"ur Theoretische Physik, Universit\"at Heidelberg,
Philosophenweg 19, D-69120 Heidelberg, Deutschland}
\author{R. Gatto}
\address{D\'epartement de Physique Th\'eorique, Universit\'e de Gen\`eve,
24 quai E.-Ansermet, CH-1211 Gen\`eve 4, Suisse}
\author{G. Nardulli and A. D. Polosa}
\address{Dipartimento di Fisica, Universit\`a di Bari and INFN Bari,
via Amendola 173, I-70126 Bari, Italia}
\date{November 1998}
\maketitle
\begin{abstract}
We evaluate the form factors governing the exclusive decays
$B\to\rho\ell\nu$, $B\to a_1 \ell\nu$, by using an effective quark-meson
lagrangian. The model  is based on meson-quark interactions, and the 
computation of the mesonic transition amplitudes is performed by
considering diagrams with heavy mesons attached to loops containing heavy 
and light constituent quarks. This approach
was successfully employed to compute the 
Isgur-Wise form factors and other hadronic observables for negative and 
positive parity heavy mesons and is presently used for exclusive 
heavy-to-light weak transitions. We also evaluate a few strong coupling 
constants appearing in chiral effective lagrangians for heavy and light mesons.
\end{abstract}

\section{Introduction}
The impressive experimental programme for the study of $B$ decays
carried out in recent years has improved our knowledge of the 
Cabibbo-Kobayashi-Maskawa 
(CKM) matrix and  CP violations. In the next few years still more abundant
data is to come, especially from the dedicated B-factories Belle \cite{belle}
and BaBar \cite{babar}.

One of the most important goals in these investigations will be a more precise
determination of the CKM matrix element  $V_{ub}$ and, to this end, 
both exclusive and inclusive $b \to u$ semileptonic transitions will be used. 
The two methods have their own 
uncertainties. Using the inclusive reaction implies the need to use perturbative QCD
methods in the region near the end-point of the lepton spectrum, where many 
resonances are present and perturbative methods are less reliable. 
This difficulty can be avoided by considering exclusive channels and
summing them up or taking them separately; however the
use of the exclusive channels forces to  evaluate the hadronic matrix elements 
by  non perturbative methods that are either  approximate or model-dependent. 
Examples of these approximations are given by non perturbative methods 
derived from Quantum-Chromo-Dynamics (QCD) first principles, i.e. 
Lattice QCD and QCD Sum rules. The drawback of
these methods is the difficulty to improve the precision and
to evaluate reliably the theoretical error, which follows from the nature of the 
approximations, i.e. the quenched approximation in Lattice QCD and a truncated 
Operator Product Expansion in QCD Sum rules. Although less fundamental, 
other approaches can be reliably used to give estimates of the hadronic 
matrix elements that appear in exclusive $b\to u$ transitions and we refer here to
constituent quark models. At the present stage of our understanding of the hadronic
interactions from first principles, they offer in our opinion a viable alternative, and 
the model dependence, which is obvious in this approach, can be used to estimate the 
overall theoretical uncertainty.  In this paper we shall follow this route 
and use a Constituent Quark Meson (CQM) model, introduced and defined 
in \cite{gatto} (hereafter referred to 
as I) to compute two semileptonic exclusive heavy to light decays, viz. 
$B$ to the light 
vector mesons $\rho$ and $a_1$. The first decay has been recently observed by the   CLEO 
collaboration \cite{cleo} (see also \cite{pdg}) that  has measured the branching 
ratio of the semileptonic decay $B \to \rho \ell \nu$:
\begin{equation}
{\cal B}(B^0 \to \rho^- \ell^+ \nu)=(2.5 \pm 0.4^{+0.5}_{-0.7}\pm0.5) \;
\times \; 10^{-4} \;\;\;. \label{cleo}
\end{equation}
This  decay will be used as a test of the
model of Ref.\cite{gatto}, since we do not introduce here any new parameter. 
On the other hand, $B \to a_1 \ell \nu$ is a prediction of this model yet to 
be tested by the experiment.

The CQM model developed in I tries to conjugate the simplicity of an effective 
lagrangian encoding the symmetries of the problem together with some dynamical 
information coming from the observed facts of confinement and
chiral symmetry breaking. In spite of the simple way in which the
dynamics is implemented, the results found in I are encouraging. We 
discussed there the following issues: leptonic constants for heavy mesons, 
Isgur-Wise form factors for heavy mesons of both negative and positive 
parity, strong coupling constants of heavy mesons and pions, radiative 
decays of $D^*,~B^*$; the comparison with data was 
satisfactory whenever it was available. 

The plan of the present paper is as follows. The model is briefly reviewed 
in Section 2. In Section 3 we compute the direct contribution to the form 
factors for
the $B \to \rho \ell \nu$, $B \to  a_1 \ell \nu$ decays, i.e. the contribution 
arising from diagrams where the weak current directly
interacts  with the quarks belonging to heavy and light mesons. 
In Section~4 we compute the strong coupling constants of $\rho$ and $ a_1$ 
to heavy mesons: these couplings are relevant for the calculation of 
the polar diagrams, i.e. the diagrams where the weak current
couples to $B$ and $\rho$ (or $a_1$) {\it via} an intermediate 
particle. These contributions (called polar contributions) 
are described in Section 5. In Section 6 we present our results 
and compare them with other approaches and with available data.
In Section 7 we draw our conclusions and in the Appendix we collect 
formulas and integrals used in the calculations. 

\section{The CQM model}

We begin with a short summary of the CQM model; for a more detailed treatment see I. The model
is an effective field theory containing a quark-meson lagrangian: 
\begin{equation}
{\cal L} ~=~{\cal L}_{\ell \ell}~+~{\cal L}_{h \ell}~.\label{lagra}
\end{equation}
The first term involves only the light degrees of freedom (a constituent quark
model for light quarks and mesons was originally suggested by Georgi and 
Manohar \cite{manohar}). To the fields considered
in I, i.e. the light quark field $\chi$ and the pseudo-scalar
$SU(3)$ octet of mesons $\pi$, we add the vector meson and axial 
vector meson octets $\rho_\mu$ and
$a_\mu$. Considering only the kinetic part of the light quarks and mesons as well as the
quark-meson interactions at the lowest order, we have for ${\cal L}_{\ell \ell} $:
\begin{eqnarray}
{\cal L}_{\ell \ell}&=&{f_{\pi}^2\over 8} \partial_{\mu} \Sigma^{\dagger} \partial^{\mu}
\Sigma +\frac{1}{2 g_V^2} tr[{\cal F(\rho)}_{\mu \nu} {\cal F(\rho)}^{\mu \nu}] 
+\frac{1}{2 g_A^2} tr[{\cal F}(a)_{\mu \nu} {\cal F}(a)^{\mu \nu}] 
\nonumber\\
&+&
{\bar \chi} (i  D^\mu \gamma_\mu -m) \chi \nonumber\\
 &+& {\bar \chi} (h_\pi
{\cal A}^\mu \gamma_\mu \gamma_5- i h_\rho \rho^{\mu} \gamma_\mu - i h_a a^{\mu} 
\gamma_\mu \gamma_5) \chi.\label{ll}
\end{eqnarray}
Let us discuss the various terms in this equation. The first three terms refer to pions,
light vector and axial vector respectively. We have $\xi=\exp(i\pi/f_\pi )$, 
$\Sigma=\xi^2$, $f_\pi=132$ MeV; the $\rho$ and $a_1$ field strengths are given by
\be
{\cal F}(x)_{\mu \nu} = \partial_\mu x_\nu - \partial_\nu x_\mu +
[x_\mu,x_\nu]~,
\ee  where, consistently with the notations of \cite{rep} (see also \cite{bando}), we 
write
\begin{equation}
\rho_\mu = i \frac{g_V}{\sqrt{2}} {\hat \rho}_\mu , ~~~~~~~~~~~~g_V = 
\frac{m_\rho}{f_\pi} \simeq 5.8~. 
\end{equation}
By analogy we also write ($m_a\simeq 1.26$ GeV is  axial vector meson mass):
\begin{equation}
a_\mu = i \frac{g_A}{\sqrt{2}} {\hat a}_\mu,
 ~~~~~~~~~~~~g_A = \frac{m_a}{f_\pi} \simeq 9.5~. \label{GA}
\end{equation}
Here $\hat \rho$, $\hat a$ are hermitean $3\times 3$ matrices
of the negative and positive parity light vector mesons.
The fourth term in Eq.~(\ref{ll}) contains
the light quarks, with $D_\mu = \partial_\mu-i {\cal V}_\mu$ and
\be
{\cal V}^\mu = {1\over 2} (\xi^\dagger \partial^\mu \xi +\xi \partial^\mu
\xi^\dagger)~. 
\ee
Therefore it gives both the kinetic term of the light quark and its coupling 
to an even number of pions. For $m$, in I we took the value $m=300$ MeV 
(for non-strange quarks).
The last three terms describe further interactions between light quarks and 
light mesons. The coupling of the light quark to an odd number of pions is 
mediated by
\be
{\cal A}^\mu = {i\over 2} (\xi^\dagger \partial^\mu \xi -\xi \partial^\mu
\xi^\dagger)~. \label{av}
\ee
Moreover, consistently with a low energy theorem for pions, 
we put $h_\pi=1$. Concerning the interactions of vector particles, we put 
$\displaystyle{h_\rho=\frac{\sqrt{2}
m^2_\rho}{g_V f_\rho},~h_a=\frac{\sqrt{2} m^2_a}{g_A f_a}}$, 
where $f_{\rho}$ and
$f_a$ are the leptonic constants. For the $\rho$ leptonic constant 
we use $f_{\rho}=0.152~{\mathrm GeV}^2$, as given by $\rho^0$, $\omega$ decay into 
$e^+ e^-$. For $f_a$ a phenomenological determination using 
$\tau \to \nu_\tau \pi \pi \pi$ was obtained in \cite{reader}, i.e. 
$f_a= 0.25 \pm 0.02~{\mathrm GeV}^2$, a result which agrees with the one found by 
QCD sum rules \cite{shu}. On the other hand from lattice QCD one obtains
$f_a=0.30 \pm 0.03 {\mathrm GeV}^2$ \cite{fa}. Since $1/f_a$ multiplies all the amplitudes 
involving the $a_1$ meson, the uncertainty on $f_a$ will induce a normalization
uncertainty on all the amplitudes involving the light axial-vector meson.
We note that our choice for $h_\rho$ and $h_a$ implements the 
hypothesis of the Vector and Axial-Vector Meson Dominance. Numerically we find:
\be 
h_\rho\simeq h_a\simeq 0.95~.
\ee
We also observe that our choice for the normalization of the light axial 
vector meson field Eq.(\ref{GA}) is conventional since $g_A$ disappears
from the physical quantities (in \cite{bando} $g_A=g_V$ is assumed). We also 
differ from the phenomenological analyses of Ref. \cite{bando} since we 
do not assume the current algebra relations $m_a^2= 2 m_\rho^2$ and 
$f_a=f_\rho$ that seem to have substantial violations.

Let us now discuss ${\cal L}_{h \ell}$, i.e. the part of the lagrangian that 
contains both  light and  heavy degrees of freedom, in particular
the heavy quark ($Q$) and mesons ($Q\bar q$). According 
to the Heavy Quark Effective Theory (HQET) \cite{neurev}, in the limit 
$m_Q \to \infty$, these mesons can be organized in spin-parity multiplets. 
We shall consider here
the negative parity spin doublet $(P,P^*)$ (e.g. $B$ and $B^*$)
and its  extension to $P$-waves, i.e. the doublet
containing the $0^+$ and  $1^+$ degenerate states  $(P_0,P_1^{*\prime})$. 
Incidentally,
we note that HQET predicts another low-lying multiplet, comprising two 
degenerate states
with $1^+$ and
$2^+$ \cite{falk}, which is of no interest here. 
In matrix notations these fields can be represented by two $4 \times 4$ Dirac 
matrices $H$ and $S$, with one spinor index for the heavy quark
and the other for the light degrees of freedom.
An explicit matrix representation is, for negative parity states:
\begin{equation}
H = \frac{(1+\slash v)}{2}\; [P_{\mu}^*\gamma^\mu - P \gamma_5 ]\\
\end{equation}
$({\bar H} = \gamma_0 H^{\dagger} \gamma_0)$, whereas, for positive 
parity states:
\begin{equation}
S={{1+\slash v}\over 2}[P_{1\mu}^{*\prime} \gamma^\mu\gamma_5-P_0]~.
\end{equation}
In these equations $v$ is the heavy meson velocity, $v^\mu P^*_{\mu}=v^\mu P_{1
\mu}^{*\prime}= 0$; $P^{*\mu}$, $P$, $P_{1\mu}^{*\prime}$ and $P_0$ are annihilation
operators normalized as follows:
\begin{eqnarray}
\langle 0|P| Q{\bar q} (0^-)\rangle & =&\sqrt{M_H}\\
\langle 0|{P^*}^\mu| Q{\bar q} (1^-)\rangle & = & \epsilon^{\mu}\sqrt{M_H}~~,
\end{eqnarray}
with similar equations for the positive parity states ($M_H=M_P=M_{P^*}$
is the common mass in the $H$ multiplet). With these notations
the heavy-light interaction lagrangian is written as follows:
\begin{eqnarray}
{\cal L}_{h \ell}&=&{\bar Q}_v i v\cdot \partial Q_v
-\left( {\bar \chi}({\bar H}+{\bar S})Q_v +h.c.\right)\nonumber \\
&+&\frac{1}{2 G_3} {\mathrm {Tr}}[({\bar H}+{\bar S})(H-S)]~,
\label{qh1}
\end{eqnarray}
where $Q_v$ is the effective heavy quark field of HQET and we have assumed 
that the fields $H$ and $S$ have the same coupling to
the quarks, which is a dynamical assumption based on a simplicity criterion 
(in I we
justify it on the basis of a four quark Nambu-Jona Lasinio interaction by 
partial 
bosonization \cite{ebert}).
After renormalization of the heavy fields $H$ and $S$
\cite{gatto} one obtains the kinetic part of the heavy meson 
lagrangian in a form that is standard for heavy meson effective chiral theories 
\cite{rep}:
\begin{eqnarray}
{\cal L}_{h \ell}&=& {\mathrm {Tr}} {\bar {\hat H}}(i v\cdot \partial
-\Delta_H){\hat H}
+{\mathrm {Tr}} {\bar {\hat S}}(i v\cdot\partial -\Delta_S) {\hat S}~.
\end{eqnarray}
Here $\Delta_H$ and $\Delta_S$ are the mass difference between the meson and the heavy 
quark at the lowest order; typical values considered in I are
$\Delta_H=0.3-0.5$ GeV. $\Delta_H$ and $\Delta_S$ are related:
for example, for  $\Delta_H = 0.4$ GeV one obtains  the value 
$\Delta_S = 0.590$ GeV \cite{gatto}.
These values correspond to a value for the $S-$multiplet mass $m=2165 \pm 50$ 
MeV; these states, called in the literature (for the charmed case) 
$D_0,D^{*\prime}_1$, have not been
observed yet, since they are expected to be rather broad.
${\hat H}$ and ${\hat S}$ are the renormalized fields 
and are given in terms of the bare 
fields $H,S$ by
\begin{eqnarray}
{\hat H} &=&  \frac{H}{\sqrt {Z_H}} \\
{\hat S} &=&  \frac{S}{\sqrt {Z_S}}. 
\end{eqnarray} 
$Z_H,~Z_S$ are  renormalization constants that have been computed in 
\cite{gatto} 
with the results (the integral $I_3$ can be found in the Appendix):
\begin{eqnarray}
Z^{-1}_H&=& \left[ (\Delta_H +m ) \frac{\partial I_3(\Delta_H)}
{\partial \Delta_H}+
I_3(\Delta_H) \right] \\ 
Z^{-1}_S&=&\left[ (\Delta_S -m ) \frac{\partial I_3(\Delta_S)}
{\partial \Delta_S}+
I_3(\Delta_S) \right]~,
\end{eqnarray}
where $m$ is the constituent light quark mass.
 
Let us finally discuss the way to compute the quark-meson 
loops arising from the previous lagrangian. As we have seen, 
the CQM model describes the interactions in terms of effective 
vertices between a
light quark, a heavy quark and a heavy meson (Eq.~(\ref{qh1}). 
We describe the heavy
quarks and heavy mesons consistently with HQET; for example 
the heavy quark propagator is given by 
\begin{equation}
{i\over  {v\cdot k + \Delta}}~,
\end{equation}
where $\Delta$ is the difference between the heavy meson and heavy quark mass and $k$ 
is the residual momentum arising from the interaction with
the light degrees of freedom.  

The light quark momentum is equal to the integrated 
loop momentum. It is therefore natural to assume an 
ultraviolet cut-off on the loop momentum of the order of the scale at which the
chiral symmetry is broken, i.e. $\Lambda \simeq 1$ GeV (in I we assumed the 
value $\Lambda=1.25$ GeV). Since the residual momentum of the heavy quark 
does not exceed few units of $\Lambda_{QCD}$ in the effective theory, 
imposing such a cut-off
does not cut any significant range of ultraviolet frequencies. We also observe
that the value of the ultraviolet cut-off $\Lambda$ is independent of 
the heavy quark mass, since it does not appear in the effective lagrangian.  

Concerning the infrared behavior, the model is not confining and thus
its range of validity cannot be extended below energies of the order of
$\Lambda_{QCD}$. In order to drop the unknown confinement part of the quark 
interaction one introduces an infrared cut-off $\mu$. These parameters 
appear in the regularized amplitudes; as discussed in 
\cite{gatto} (see also \cite{ebert}) we have  chosen a proper 
time regularization; the regularized form for the light quark propagator 
(including integration over momenta) is
\begin{equation}
\int d^4 k_E  \frac{1}{k_E^2+m^2} \to \int d^4 k_E \int_{1/
\Lambda^2}^{1/\mu^2} ds\; e^{-s (k_E^2+m^2)}~, \label{cutoff}
\end{equation}
where $\mu$ and $\Lambda$ are
the infrared and  ultraviolet cut-offs. For $\mu$ in I we assumed the value:
$\mu=300$ MeV. For a different choice of the 
cut-off prescription in related models see \cite{holdom}, \cite{bardeen}.

\section{ $B \to \rho$ and $B \to a_1$ form factors: evaluation of the direct 
contributions.}

The  form factors for the semileptonic decays $B\to\rho \ell\nu$ 
and $B\to a_1 \ell\nu$ can be written as follows ($q=p-p^\prime$):
\begin{eqnarray}
<\rho^+(\epsilon(\lambda),p^\prime)|\overline{u}\gamma_\mu 
(1-\gamma_5)b|\bar{B^0}(p)>
& = & \frac{2 V(q^2)}{m_B+m_{\rho}}\epsilon_{\mu\nu\alpha\beta}
\epsilon^{*\nu}p^\alpha p^{\prime\beta}\nonumber\\
&-& i \epsilon^{*}_{\mu}(m_{B} + m_{\rho})  A_{1}(q^{2})
\nonumber\\
&+& i (\epsilon^{*}\cdot q)
\frac{(p + p^\prime)_{\mu}}{m_B +  m_{\rho}}  A_{2}(q^{2})
\\
&+& i  (\epsilon^{*}\cdot  q)
\frac{2  m_{\rho}}{q^{2}} q_{\mu} [A_{3}(q^{2})  - A_{0}(q^{2})]
\nonumber\;\; ,
\end{eqnarray}
where
\begin{equation}
A_{3}(q^{2})  = \frac{m_{B} + m_{\rho}}{2  m_{\rho}} A_{1}(q^{2})
- \frac{m_{B} - m_{\rho}}{2  m_{\rho}} A_{2}(q^{2}) \;\; ,
\end{equation}
and

\begin{eqnarray}
<a^{+}_1(\epsilon(\lambda),p^\prime)|\overline{q^\prime}
\gamma_\mu(1-\gamma_5)b|B(p)>
& = & \frac{2 A(q^2)}{m_B+m_a}\epsilon_{\mu\nu\alpha\beta}
\epsilon^{*\nu}p^\alpha p^{\prime\beta}\nonumber\\
&-& i \epsilon^{*}_{\mu}(m_{B} + m_{a})  V_{1}(q^{2})
\nonumber\\
&+& i (\epsilon^{*}\cdot q)
\frac{(p + p^\prime)_{\mu}}{m_B +  m_a}  V_{2}(q^{2})
\\
&+& i  (\epsilon^{*}\cdot  q)
\frac{2  m_a}{q^{2}} q_{\mu} [V_{3}(q^{2})  - V_{0}(q^{2})]
\nonumber\;\; ,
\end{eqnarray}
where $m_a $ is the $a_1$ mass and 
\begin{equation}
V_{3}(q^{2})  = \frac{m_{B} + m_{a}}{2  m_{a}} V_{1}(q^{2})
- \frac{m_{B} - m_{a}}{2  m_{a}} V_{2}(q^{2}) \;\; ,
\end{equation}
We note that, having used this parameterization for
the weak matrix elements\cite{bsw}, at $q^2=0$ the following conditions hold
\begin{eqnarray}
A_{3}(0)  &=& A_{0}(0)\label{A3}\\
V_{3}(0)  &=& V_{0}(0)\label{V3}\;.
\end{eqnarray}
\par
The contribution we consider in this section arises from  
diagrams where the weak current couples directly to the quarks belonging to
the light and heavy mesons (see Fig. 1).

These diagrams are computed using the rules described in the previous Section.
The results of a straightforward, but lengthy calculation are as follows:
\begin{eqnarray}
V^{D}(q^{2}) &=& -\frac{m_{\rho}^2}{f_{\rho}} \sqrt{\frac{Z_H} {m_B}}
\left( \Omega_1 - m Z \right) (m_B + m_{\rho})\\
A^{D}_1 (q^{2}) &=& \frac{2 m_{\rho}^2}{f_{\rho}} \sqrt{Z_H m_B} 
\frac{1}{m_B + m_{\rho}}
\nonumber \\ 
&& \left[ (m^2 + m m_{\rho} {\bar{\omega}}) Z  -{\bar{\omega}} 
m_{\rho}\Omega_1 - m_\rho 
\Omega_2 -2 \Omega_3 -\nonumber \right.\\
&& \left. \Omega_4 -\Omega_5 -2 {\bar{\omega}} \Omega_6 \right]\\
A^{D}_2(q^{2}) &=& \frac{m_{\rho}^2}{f_\rho }\sqrt{\frac{Z_H}{m_B}} 
\left( m Z -\Omega_1 - 2 \frac{\Omega_6}{m_\rho} \right) (m_B + m_{\rho})\\
A^{D}_0 (q^{2}) &=& -\frac{m_\rho }{f_\rho} \sqrt{Z_H m_B} 
\left[\Omega_1 \left(  m_\rho {\bar{\omega}} +2 m \frac{q^2}{m_B^2} -
 \frac{r_1}{m_B}\right) +  m_\rho \Omega_2 + \nonumber \right.\\
&&\left. 2\Omega_3 + \Omega_4 (1- 
2 \frac{q^2}{m_B^2}) + \Omega_5 + 2\Omega_6 \left(  \bar{\omega}- \frac
{r_1}{m_B m_\rho} \right)-\nonumber\right.\\
&&\left.  Z(m^2  - m \frac{r_1}{m_B}  +  m m_\rho {\bar{\omega}})
\right]
\end{eqnarray}
where
\begin{equation}
{\bar{\omega}}=\frac{m_B^2+m_\rho^2-q^2}{2 m_B m_\rho}~,
\end{equation}
and
\begin{equation}
r_1=\frac{m_B^2-q^2-m^2_\rho}{2}\label{r2}
\end{equation}
and the functions $Z$, $\Omega_j$ are given by the formulae of the Appendix
with $\Delta_1=\Delta_H$, $\Delta_2=\Delta_1 -m_\rho {\bar{\omega}}$, 
$x=m_\rho$; $m$ is the constituent light quark mass.

\newpage
The calculation  for the $ B \to a_1$ transition is similar. The results are:
\begin{eqnarray}
A^{D}(q^2) &=&- \frac{m_{a}^2}{f_{a}} \sqrt{\frac{Z_H} {m_B}}
\left( \Omega_1 - m Z -\frac{2m}{m_a} \Omega_2 \right) (m_B + m_a) \\
V^{D}_1 (q^{2})&=&\frac{2 m_{a}^2}{f_a} \sqrt{Z_H m_B} 
\frac{1}{m_B + m_a}
\nonumber \\ 
&&\left[(-m^2 + m m_a \bar{\omega})Z + 2m\Omega_1 - \bar{\omega}m_a\Omega_1+
\nonumber \right.\\
&& \left.+(2m\bar{\omega}  
-m_a )\Omega_2-2\Omega_3 -\Omega_4-\Omega_5
-2\bar{\omega}\Omega_6  \right]\\
V^{D}_2(q^{2}) &=& \frac{m_a^2}{f_a}\sqrt{\frac{Z_H}{m_B}} 
\left( m Z -\Omega_1 - 2 \frac{\Omega_6}{m_a}+2\frac{m}{m_a}\Omega_2 
\right) (m_B + m_a)\\
V^{D}_0 (q^{2}) &=& -\frac{m_a }{f_a} \sqrt{Z_H m_B} 
\left[\Omega_1 \left( m_a \bar{\omega}+2m \frac{q^2}{m^{2}_B}-  
\frac{r_1^{\prime}}{m_B}
-2m\right) + \nonumber \right.\\
&& \left. \Omega_2 \left( m_a + 2m \frac{ r_1^\prime}{m_B m_a} -
 2 m \bar{\omega}\right)+
2\Omega_3 + \Omega_4 \left(1- 2\frac{q^2}{m^{2}_B} \right)+\Omega_5+ 
\nonumber
\right.\\
&&\left. 2\Omega_6 \left(\bar{\omega} -   \frac{r_1^\prime}{m_B m_a} \right)+ 
Z\left(m^2+m\frac{r_1^\prime}{m_B}- m m_a \bar{\omega} \right)\right] ~,
\end{eqnarray}
where now:
\begin{eqnarray}
{\bar{\omega}}&=&\frac{m_B^2+m_a^2-q^2}{2 m_B m_a}\\
r_1^\prime & = &\frac{m_B^2-q^2-m^2_a}{2}~.
\label{r2p}
\end{eqnarray}
The previous results for the form factors can be used directly for the
numerical analysis. In order to allow an easier way of using our results
we have fitted these formulas by the simple parameterization:
\begin{equation}
F^D(q^2)=\frac{F^D(0)}{1~-~a_F \left(\dd\frac{q^2}{m_B^2}\right) +~b_F 
\left(\dd\frac{q^2}{m_B^2}\right)^2}
\label{16b}
\end{equation}
\noindent
for a generic form factor $F^D(q^2)$; $a_F~,b_F~$ have been fitted by a 
numerical analysis performed up to $q^2=16~{\mathrm GeV}^2$, both for $\rho$
and $a_1$ mesons. We have collected the fitted values in Table \ref{t:tab1}.
We note explicitly that the results for $B \to a_1$ form factors at $q^2=0$
are proportional to the factor 
$(0.25~{\mathrm GeV}^2/f_a)$. Besides the normalization
uncertainty due to $f_a$, we estimate a theoretical error of $15\%$ on these
parameters.

\section{Strong coupling constants}

In this Section we compute the strong couplings $HH\rho$, $H S\rho$, $HH a_1$, 
$H S a_1$. As discussed in the introduction they are relevant for the
calculation of the polar contribution to the form factors. 
We parameterize these couplings by considering
the following effective Lagrangians (we follow the notations introduced in 
\cite{rep}):
\begin{eqnarray}
{\cal L}_{H H \rho}&=&i\lambda {\mathrm {Tr}}(\overline{H} H \sigma^{\mu \nu}
{\cal F(\rho)}_{\mu \nu})  -i \beta {\mathrm {Tr}}(\overline{H} H \gamma^\mu 
\rho_\mu)
\\ \label{strong1}
{\cal L}_{H S \rho}&=&-i \zeta {\mathrm {Tr}}(\overline{S} 
H \gamma^\mu \rho_\mu)+
i\mu {\mathrm {Tr}}(\overline{S} H \sigma^{\mu \nu}
{\cal F(\rho)}_{\mu \nu})
\\ \label{strong2}
{\cal L}_{H H a_1}&=& -i \zeta_A {\mathrm {Tr}}(\overline{H} H 
\gamma^\mu a_\mu )+
i\mu_A {\mathrm {Tr}}(\overline{H} H \sigma^{\mu \nu}
{\cal F}(a)_{\mu \nu} )
\\ \label{strong3}
{\cal L}_{H S a_1}&=&i\lambda_A {\mathrm {Tr}}(\overline{S} H \sigma^{\mu \nu}
{\cal F}(a)_{\mu \nu} )  -i \beta_A {\mathrm {Tr}}(\overline{S} H \gamma^\mu 
a_\mu )~.
\label{strong4}
\end{eqnarray}
The strong couplings can be computed by matching the effective meson 
lagrangian of Eqs.(\ref{strong1})-(\ref{strong4}) with the quark-meson 
lagrangian (\ref{qh1}), i.e. considering triangular quark loops 
with external legs representing light and heavy mesons. The calculation
is similar to the one of the previous section. The results are 
as follows:
\begin{eqnarray}
\lambda &=&\frac{m^2_{\rho}}{ \sqrt{2} g_V f_{\rho}} Z_H (-\Omega_1+ m Z)\\
\beta &=&\sqrt{2}\frac{m^2_{\rho}}{ g_V f_{\rho}}   Z_H [2 m \Omega_1+
 m_\rho \Omega_2 + 2 \Omega_3 - \Omega_4 + \Omega_5- 
 m^2 Z ]~.
\end{eqnarray}
Here the functions $Z$, $\Omega_j$ are given by the formulae of the Appendix
with $\Delta_1=\Delta_H$, $x=m_\rho$, $\omega=m_\rho/(2m_B)$ (we keep here 
the first $1/m_Q$ correction). Moreover
\begin{eqnarray}
\mu &=& \frac{m^2_\rho}{\sqrt{2} g_V f_{\rho}} \sqrt{Z_H Z_S}( -\Omega_1-
2 \frac{\Omega_6}{m_\rho}+ m Z) \\
\zeta &=& \frac{\sqrt{2} m^2_\rho}{g_V f_{\rho}} \sqrt{Z_H Z_S} 
\left(m_\rho  \Omega_2 +2 \Omega_3 +\Omega_4 +\Omega_5 -m^2 Z \right)~.
\end{eqnarray}
Here the functions $Z$, $\Omega_j$ are given by the formulae of the Appendix
with $\Delta_1=\Delta_H$, $\Delta_2=\Delta_S$, $x=m_\rho$ and 
$\omega=(\Delta_1-\Delta_2)/{m_\rho}$.
For the axial-vector $a_1$ couplings to $H$ and $S$ states we find 
\begin{eqnarray}
\lambda_A &= & \frac{m^2_a}{\sqrt{2} g_A f_a}   \sqrt{Z_H Z_S} 
( -\Omega_1 +2\Omega_2\frac{m}{m_a} + m Z)\\
\beta_A & = & \sqrt{2}\frac{m^2_a}{ g_A f_a}   \sqrt{Z_H Z_S} 
(m_a \Omega_2 +2\Omega_3   -\Omega_4 + \Omega_5 
  + m^2 Z)~,
\end{eqnarray}
where $Z$, $\Omega_j$ are given by the formulae of the Appendix
with $\Delta_1=\Delta_H$, $\Delta_2=\Delta_S$, $x=m_a$ and 
$\omega=(\Delta_1-\Delta_2)/{m_a}$. Moreover
\begin{eqnarray}
\mu_A & = & \frac{m^2_a}{\sqrt{2} g_A f_a} Z_H \left( m \left(Z+  
2\frac{\Omega_2}{m_a}\right) 
- \Omega_1 -2\frac{\Omega_6}{m_a}\right) \\
\zeta_A & = & \frac{\sqrt{2} m^2_a}{g_A f_a} Z_H
( -2m\Omega_1+ m_a\Omega_2 + 2\Omega_3+\Omega_4+\Omega_5+m^2 Z) ~,
\end{eqnarray} 
where $\Delta_1=\Delta_H$, $x=m_a$, $\omega=m_a/(2m_B)$ 
Numerically we get the following results:
$$
\begin{array}{cclcccl}
\lambda &=& 0.60~{\mathrm {GeV}}^{-1}
&\hspace{0.5truecm}&  \lambda_A &=& 0.85 \times
(0.25~ {\mathrm GeV}^2/f_a)~{\mathrm {GeV}}^{-1}\nonumber \\
\beta &=& -0.86
& & \beta_A &=& -0.81 \times (0.25~ {\mathrm GeV}^2/f_a) \nonumber \\
\mu &=& 0.16~{\mathrm {GeV}}^{-1}
& &  \mu_A &=& 0.23 \times (0.25~ {\mathrm GeV}^2/f_a)~{\mathrm {GeV}}^{-1} \\
\zeta &=& 0.01
& &\zeta_A &=&  0.15 \times (0.25~ {\mathrm GeV}^2/f_a)~.
\end{array}
$$
A discussion about the theoretical uncertainties of these results is in order.
We have explicitly written down the dependence on $f_a$ of the strong 
coupling constants involving the light axial-vector meson, since as noted
before, this is a major source of theoretical uncertainty for these constants.
Another source of spreading in the reported values is the variation of 
$\Delta_H$ in the range $\Delta_H=0.3-0.5$ GeV (we use $\Delta_H=0.4$ GeV
in the calculation). This produces a significant uncertainty
only for $\zeta,\beta_A,\zeta_A$ since we obtain $\zeta = 0.01\pm 0.19$,
$\beta_A = -0.81^{+0.45}_{-0.24}$ and $\zeta_A =  0.15^{+0.16}_{-0.14}$ while
in the other cases only a few percent variation is observed.
For the other constants: $\lambda$, $\mu$, $\lambda_A$, $\mu_A$, a
theoretical uncertainty of $\pm 15 \%$ can be guessed. This estimate follows
for example from a different evaluation of the $\lambda$ parameter performed
in I. For other determinations of the coupling constant $\lambda$ see 
\cite{aliev} (QCD sum rules and light cone sum rules) and \cite{defazio2}
(this paper uses data from $D^*$ decays together with vector meson dominance).

\section{ $B \to \rho$ and $B \to a_1$ form factors: evaluation of 
the polar contributions.}

The polar contributions are given by  diagrams where the weak 
current is coupled to $B$ and to the light vector or axial vector meson 
by an intermediate heavy meson state (see Fig. 2). These diagrams, 
because of the heavy
meson propagator, produce a typical polar behavior of the form 
factors, in the form
\begin{equation}
F^P(q^2)=\frac{F^P(0)}{1~-~ \frac{q^2}{m_P^2}}~.
\label{fp}
\end{equation}
\noindent
This behaviour is certainly valid near the pole; we assume its validity for 
the whole $q^2$ range, which can be justified on the basis of the minor 
numerical role of this polar
contribution, as compared to the direct one, in the region of low $q^2$,
at least for the form factors $A_1^P, A_2^P$ (see the
numerical results at the end of this Section). The assumption 
(\ref{fp}) cannot be made
for the form factors $A_0^P(q^2)$ and
$V_0^P(q^2)$, as we discuss below, and is also less reliable for 
$A^P(q^2)$ and $V^P(q^2)$ (see Table \ref{t:tabp}).

The values at $q^2=0$ in Eq.~(\ref{fp}) can be easily computed in term of 
the strong coupling constants defined in the previous Section and using 
the leptonic decay constants ${\hat F}$ and ${\hat F}^+$ that give the 
coupling of the intermediate states to the currents.
Neglecting logarithmic corrections, ${\hat F}$ and ${\hat F}^+$ are related,
in the infinite heavy quark mass limit, to the
leptonic decay constant $f_B$ and $f^+$ defined by
\begin{eqnarray}
\langle 0|{\bar q} \gamma^\mu \gamma_5 b  |B(p)\rangle
&=& i p^\mu f_B \\
\langle 0|{\bar q} \gamma^\mu  b  |B_0(p)\rangle
&=&  p^\mu f^+ ~,
\end{eqnarray}
by the relations $f_B={\hat F}/\sqrt{m_B}$
and $f^+={\hat F}^+/\sqrt{m_{B_0}}$ ($B_0$ is the $S$ state with $J^P=0^+$ and 
$b\bar q$ content).
These couplings have been computed in \cite{gatto} with the results
given in  Table \ref{t:tabf} for different values of the
parameters.

For the values $F^P(0)$ we obtain the following results for the $B\to \rho$ 
transition.

\begin{eqnarray}
V^P (0)&=&  -\sqrt{2} g_V \lambda {\hat F} \frac{m_B +m_\rho}{ m_B^{3/2}}\\
A^P_1 (0) &=&  \frac{\sqrt{2 m_B}g_V {\hat F}^+}{m_{B_0} (m_B+m_{\rho})}
(\zeta-2\mu {\bar \omega} m_{\rho})\\
A^P_2 (0) &=& -\sqrt{2} g_V \mu {\hat F}^+ \frac{\sqrt{m_B} (m_B+m_\rho)}
{m_{B_0}^2}~.
\end{eqnarray}
where ${\bar \omega}=m_B/(2 m_\rho)$.
For $A^P_0 (q^2)$, we have to implement the condition contained in 
Eq.~(\ref{A3}); for instance a possible choice is
\begin{equation}
A^P_0 (q^2)= A^P_3(0) + g_V \beta {\hat F} \frac{1}{m_\rho\sqrt{2 m_B}}
\frac{q^2}{m^2_B-q^2}~. 
\label{AP0}
\end{equation}
For the  $B\to a_1$ transition we have:
\begin{eqnarray}
A^P (0)&=&  -\sqrt{2} g_A \lambda_A {\hat F}^+ \frac{m_B +m_a}{ m_B^{3/2}}\\
V^P_1 (0) &=&  \frac{\sqrt{2 m_B}g_A {\hat F}}{m_B (m_B+m_a)} 
(\zeta_A-2\mu_A {\bar \omega} m_a) \\
V^P_2 (0) &=& -\sqrt{2} g_A \mu_A {\hat F} \frac{\sqrt{m_B} (m_B+m_a)}
{m_B^2}~,
\end{eqnarray}
where ${\bar \omega}=m_B/(2 m_a)$. Similarly to the previous discussion
for $A^P_0 (q^2)$ we can put for instance:
\begin{equation}
V^P_0 (q^2)= V^P_3(0) + g_A \beta_A {\hat F}^+ \frac{1}{m_a\sqrt{2 m_B}}
\frac{q^2}{m^2_{B_0}-q^2} ~.
\label{VP0}
\end{equation}
We note that Eqs.~(\ref{AP0}) and (\ref{VP0}) have been written down only 
as an example of
a possible behaviour of these form factors satisfying the given constraints. 
For massless
leptons they do not contribute to the semileptonic width and can be neglected.

Numerically we obtain the results in Table \ref{t:tabp} where we have also 
reported the values of the pole masses for all the form factors except 
$A^P_0 (q^2)$ and $V^P_0 (q^2)$ because of Eqs.~(\ref{AP0}) and (\ref{VP0}). 
Similarly to the previous analyses an overall
uncertainty of $\pm 15\%$ can be assumed. In Fig. 3 we plot the form factors
$A_1$ and $A_2$ for the semileptonic decay $B \to \rho$. In Fig. 4 are shown
the form factors $A$, $V_1$ and $V_2$ for the semileptonic decay $B \to a_1$.
Since the behaviour in Eqs.~(\ref{AP0}) and (\ref{VP0}) is only guessed, 
we have 
not included the form factors $A^P_0 (q^2)$ and $V^P_0 (q^2)$ in Fig. 3 and 4;
in addition $V(q^2)$ is not reported in Fig. 3 since our prediction is 
affected by a large error (see the discussion in the next section).
Note that the theoretical error is not included in Fig. 3 and 4; one should 
refer to the numbers in Tables \ref{t:tab1} and \ref{t:tabp}.

\section{Branching ratios and widths}
In this Section we compute the branching ratios and widths for semileptonic
decays using the numerical results for form factors reported in Table I,II.
Let us first compare our results for the $B \to \rho$ form factors with 
those obtained by other methods (see Table IV). These form factors (as well
as those concerning the transition $B \to a_1$) are obtained by adding
the direct and polar contributions:
\begin{equation}
F(q^2)=F^D(q^2) + F^P(q^2),
\end{equation}
where $F^D(q^2)$ was introduced in Section III and $F^P(q^2)$ in Section V.
Our result for the vector form factor $V^{\rho}(q^2)$ 
is affected by a large error since it arises from
the sum of two terms opposite in sign and almost equal in absolute value.
Apart from this large uncertainty, our results are in relative good 
agreement with the results of QCD sum rules, but they are in general 
higher than those obtained by other approaches.
For the $B\to \rho \ell\nu$ decay width and branching ratio
we obtain (using $V_{ub}=0.0032$, 
$\tau_B=1.56 \times 10^{-12}$ s):
\begin{eqnarray}
{\cal B}(\bar B^0 \to \rho^+ \ell \nu) &=& (2.5 \pm 0.8) \times 10^{-4} \nonumber \\
\Gamma_0(\bar B^0 \to \rho^+ \ell \nu) &=& (4.4 \pm 1.3) \times 10^{7} \; s^{-1} 
\nonumber \\
\Gamma_+ (\bar B^0 \to \rho^+ \ell \nu)&=& (7.1 \pm 4.5) \times 10^{7} \; 
s^{-1} \nonumber \\
\Gamma_- (\bar B^0 \to \rho^+ \ell \nu)&=& (5.5 \pm 3.7) \times 10^{7} \; 
s^{-1} \nonumber \\
(\Gamma_+ + \Gamma_-) (\bar B^0 \to \rho^+ \ell \nu)&=& (1.26 \pm 0.38) \times
10^8 \; s^{-1}
\end{eqnarray}
where $\Gamma_0$, $\Gamma_+$, $\Gamma_-$ refer to the $\rho$ helicities.
The branching ratio for $B\to \rho \ell\nu$ is in agreement with the 
experimental result quoted in the introduction, Eq.~(\ref{cleo}).

Let us now discuss the theoretical uncertainty of these results. The large error
of $V^\rho (0)$ affects significantly the values of $\Gamma_+$ and $\Gamma_-$, 
whose errors are correlated;
it has however no effect on $\Gamma_0$ and a small effect 
on the branching ratio, which increases at most by $8\%$. The theoretical 
uncertainties on $A_1^\rho (0)$ and $A_2^\rho (0)$ are likely to be related. 
To get the theoretical error on the widths we have added in quadrature the
error induced by $V^\rho (0)$ and a common $\pm 15\%$ error on 
$A_1^\rho (0)$ and $A_2^\rho (0)$.

Having used the decay $B\to \rho \ell\nu$ as a test of the CQM model, we can
now consider the $B\to a_1 \ell\nu $ semileptonic decay. 
The results obtained are:
\begin{eqnarray}
{\cal B}(\bar B^0 \to a_1^+ \ell \nu) &=& (8.4 \pm 1.6) \times 10^{-4} 
\nonumber \\
\Gamma_0 (\bar B^0 \to a_1^+ \ell \nu)&=& (4.0 \pm 0.7) \times 10^{8} \; 
s^{-1} \nonumber \\
\Gamma_+ (\bar B^0 \to a_1^+ \ell \nu)&=& (4.6 \pm 0.9) \times 10^{7} \; 
s^{-1} \nonumber \\
\Gamma_- (\bar B^0 \to a_1^+ \ell \nu)&=& (0.98 \pm 0.18) \times 10^{8} \; 
s^{-1} \label{ba1}
\end{eqnarray}
where $\Gamma_0$, $\Gamma_+$, $\Gamma_-$ refer to the $a_1$ helicities.
We have included in the determination of these decay widths only the 
normalization uncertainty arising from $f_a$; the lower values correspond to
$f_a=0.30~{\mathrm GeV}^2$ while the higher values to 
$f_a=0.25~{\mathrm GeV}^2$. One should also take into account the 
theoretical errors arising from the values of the form factors at $q^2=0$;
they are more difficult to estimate reliably and are not included here. 
In any case the over-all theoretical uncertainty is larger (presumably
by a factor of two) than the one reported in the previous formula.

\section{Conclusions}
The main conclusion of this paper can be read from eqs.(\ref{ba1}). 
We predict
a branching ratio for the decay $\bar B^0 \to a_1^+ \ell \nu$ significantly
larger than the branching ratio for $\bar B^0 \to \rho^+ \ell \nu$;
in spite of the theoretical uncertainties inherent to the CQM model,
which we have discussed in the previous Sections, this is a remarkable outcome.
A consequence of this result
is that the  $ B \to a_1 \ell \nu$ decay channel 
might account for around $50\%$
of the semileptonic  $ B \to X_u \ell \nu$ decay channel (evaluated, 
for example, by the parton model), whereas the $ B \to \rho \ell \nu$ 
decay channel adds another $15\%$; given the relevance of these results for the
determination of $V_{ub}$, it would be interesting to test these predictions
in the future by other theoretical methods and, hopefully, by experimental data.

\section{Appendix}
In the paper we have introduced  several integrals and parameters that we list
in this Appendix. 
\begin{eqnarray}
I_0(\Delta)&=& \frac{iN_c}{16\pi^4} \int^{\mathrm {reg}}
\frac{d^4k}{(v\cdot k + \Delta + i\epsilon)} \nonumber \\
&=&{N_c \over {16\,{{\pi }^{{3/2}}}}}
\int_{1/{{{\Lambda}^2}}}^{1/{{{\mu }^2}}} {ds \over {s^{3/2}}}
\; e^{- s( {m^2} - {{\Delta }^2} ) }
\left( {3\over {2\,s}} + {m^2} - {{{\Delta}}^2} \right)
[1+{\mathrm {erf}}(\Delta\sqrt{s})]\nonumber \\
&-& \Delta {{N_c m^2}\over {16 \pi^2}} \Gamma(-1,{{{m^2}}
\over {{{\Lambda}^2}}},{{{m^2}}\over {{{\mu }^2}}})
\\
I_1&=&\frac{iN_c}{16\pi^4} \int^{reg} \frac{d^4k}{(k^2 - m^2)}
={{N_c m^2}\over {16 \pi^2}} \Gamma(-1,{{{m^2}}
\over {{{\Lambda}^2}}},{{{m^2}}\over {{{\mu }^2}}})
\\
I_1^{\prime}&=&\frac{iN_c}{16\pi^4} \int^{\mathrm {reg}} d^4
k\frac{k^2}{(k^2 - m^2)}
={{N_c m^4}\over {8 \pi^2}} \Gamma(-2,{{{m^2}}
\over {{{\Lambda}^2}}},{{{m^2}}\over {{{\mu }^2}}})\\
I_2&=&-\frac{iN_c}{16\pi^4} \int^{\mathrm {reg}}  \frac{d^4k}{(k^2 - m^2)^2}=
\frac{N_c}{16\pi^2} \Gamma(0,\frac{m^2}{\Lambda^2}, \frac{m^2}{\mu^2})\\
I_3(\Delta) &=& - \frac{iN_c}{16\pi^4} \int^{\mathrm {reg}}
\frac{d^4k}{(k^2-m^2)(v\cdot k + \Delta + i\epsilon)}\nonumber \\
&=&{N_c \over {16\,{{\pi }^{{3/2}}}}}
\int_{1/{{\Lambda}^2}}^{1/{{\mu }^2}} {ds \over {s^{3/2}}}
\; e^{- s( {m^2} - {{\Delta }^2} ) }\;
\left( 1 + {\mathrm {erf}} (\Delta\sqrt{s}) \right)\\
I_4(\Delta)&=&\frac{iN_c}{16\pi^4}\int^{\mathrm {reg}}
\frac{d^4k}{(k^2-m^2)^2 (v\cdot k + \Delta + i\epsilon)} \nonumber\\
&=&\frac{N_c}{16\pi^{3/2}} \int_{1/\Lambda^2}^{1/\mu^2} \frac{ds}{s^{1/2}}
\; e^{-s(m^2-\Delta^2)} \; [1+{\mathrm {erf}}(\Delta\sqrt{s})]~.
\end{eqnarray}
where $\Gamma$ is the generalized incomplete gamma function and erf
is the error function. Moreover, having defined:
\begin{equation}
\sigma(x,\Delta_1,\Delta_2,\omega)={{{\Delta_1}\,\left( 1 - x \right)  +
{\Delta_2}\,x}\over {{\sqrt{1 + 2\,\left(\omega -1  \right) \,x +
2\,\left(1-\omega\right) \,{x^2}}}}}~,
\end{equation}
we have:
\begin{eqnarray}
I_5(\Delta_1,\Delta_2, \omega) & = & \frac{iN_c}{16\pi^4} \int^{\mathrm {reg}}
\frac{d^4k}{(k^2-m^2)(v\cdot k + \Delta_1 + i 
\epsilon )
(v'\cdot k + \Delta_2 + i\epsilon )} \nonumber \\
 & = & \int_{0}^{1} dx \frac{1}{1+2x^2 (1-\omega)+2x
(\omega-1)}\times\nonumber\\
&&\Big[ \frac{6}{16\pi^{3/2}}\int_{1/\Lambda^2}^{1/\mu^2} ds~\sigma
\; e^{-s(m^2-\sigma^2)} \; s^{-1/2}\; (1+ {\mathrm {erf}}
(\sigma\sqrt{s})) +\nonumber\\
&&\frac{6}{16\pi^2}\int_{1/\Lambda^2}^{1/\mu^2}
ds \; e^{-s(m^2-2\sigma^2)}\; s^{-1}\Big]~.
\end{eqnarray}
We also define, if $q^\mu= x v^{\prime\mu}$,  $\omega=v\cdot v^{\prime}$, 
$\Delta_2=\Delta_1- x ~ \omega$, the formula: 
\begin{eqnarray}
Z &=&  \frac{iN_c}{16\pi^4} \int^{\mathrm {reg}}
\frac{d^4k}{(k^2-m^2)[(k+q)^2-m^2](v\cdot k + \Delta_1 + i\epsilon)}\nonumber \\
&=&\frac{I_5(\Delta_1, x/2,\omega)-I_5(\Delta_2,- x/2,\omega)}{2 x}~.
\end{eqnarray}
We use in the text the following combinations of the previous integrals:
\begin{eqnarray}
K_1&=&m^2 Z -I_3(\Delta_2)\nonumber \\
K_2&=&\Delta_1^2 Z -\frac{I_3(x/2)-
I_3(-x/2)}{4 x}[\omega ~ x + 2 \Delta_1]
\nonumber \\
K_3&=&\frac{x^2}{4} Z +\frac{I_3(\Delta_1)-3
I_3(\Delta_2)}{4}+\frac{\omega}{4}[\Delta_1 I_3(\Delta_1)-
\Delta_2 I_3(\Delta_2)]
\nonumber \\
K_4&=&\frac{x \Delta_1}{2} Z +\frac{\Delta_1[I_3(\Delta_1)-
I_3(\Delta_2)]}{2 x}+\frac{I_3(x/2)-I_3(-x/2)}{4} \nonumber \\
\Omega_1&=&\frac{ I_3(-x/2)-I_3(x/2)+\omega[I_3(\Delta_1)-
I_3(\Delta_2)]}{2 x (1-\omega^2)} - \frac{[\Delta_1-\omega x/2]Z}
{1-\omega^2}
\nonumber \\
\Omega_2&=&\frac{ -I_3(\Delta_1)+I_3(\Delta_2)-\omega[I_3(-x/2)-I_3(x/2)]}
{2 x (1-\omega^2)} - \frac{[x/2- \Delta_1\omega ]Z}{1-\omega^2}
\nonumber \\
\Omega_3&=&\frac{K_1}{2}+\frac{2 \omega K_4-K_2-K_3}{2(1-\omega^2)}
\nonumber \\
\Omega_4&=&\frac{-K_1}{2(1-\omega^2) }+\frac{3 K_2-6 \omega K_4+K_3
(2\omega^2+1)}{2(1-\omega^2)^2}
\nonumber \\
\Omega_5&=&\frac{-K_1}{2(1-\omega^2) }+\frac{3 K_3-6 \omega K_4+K_2
(2\omega^2+1)}{2(1-\omega^2)^2}
\nonumber \\
\Omega_6&=&\frac{K_1\omega}{2(1-\omega^2) }+\frac{2 K_4(2\omega^2+1)-
3\omega( K_2+K_3)
}{2(1-\omega^2)^2}~.
\nonumber \\
\end{eqnarray}

\twocolumn
\par
\noindent
\vspace*{1cm}
\par
\noindent
{\bf Acknowledgments}
\par
\noindent
A.D. acknowledges the support of a ``Marie Curie'' TMR research fellowship
of the European Commission under contract ERBFMBICT960965 in the first stage
of this work. He was also supported by the EC-TMR (European Community 
Training and Mobility of Researchers) Program on ``Hadronic Physics with 
High Energy Electromagnetic Probes''. A.D.P. acknowledges support from
I.N.F.N. Italy. This work has also been carried out in part under the
EC program Human Capital and Mobility, contract UE
ERBCHRXCT940579 and OFES 950200.

\onecolumn

\vskip1truecm
\begin{figure}
\epsfxsize=9truecm
\centerline{\epsffile{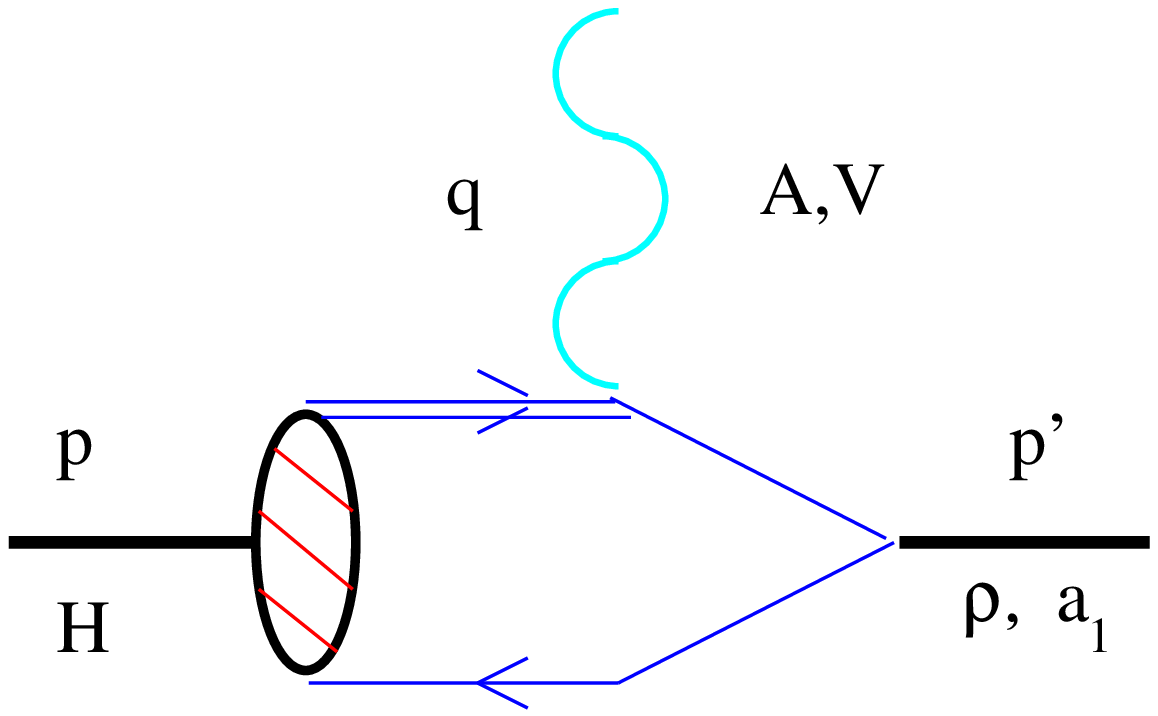}}
\noindent
{\bf Fig. 1} - {Diagram for the direct contribution to the form factor
$B \to \rho$ and $B \to a_1$.}
\end{figure}

\vskip2truecm
\begin{figure}
\epsfxsize=13truecm
\centerline{\epsffile{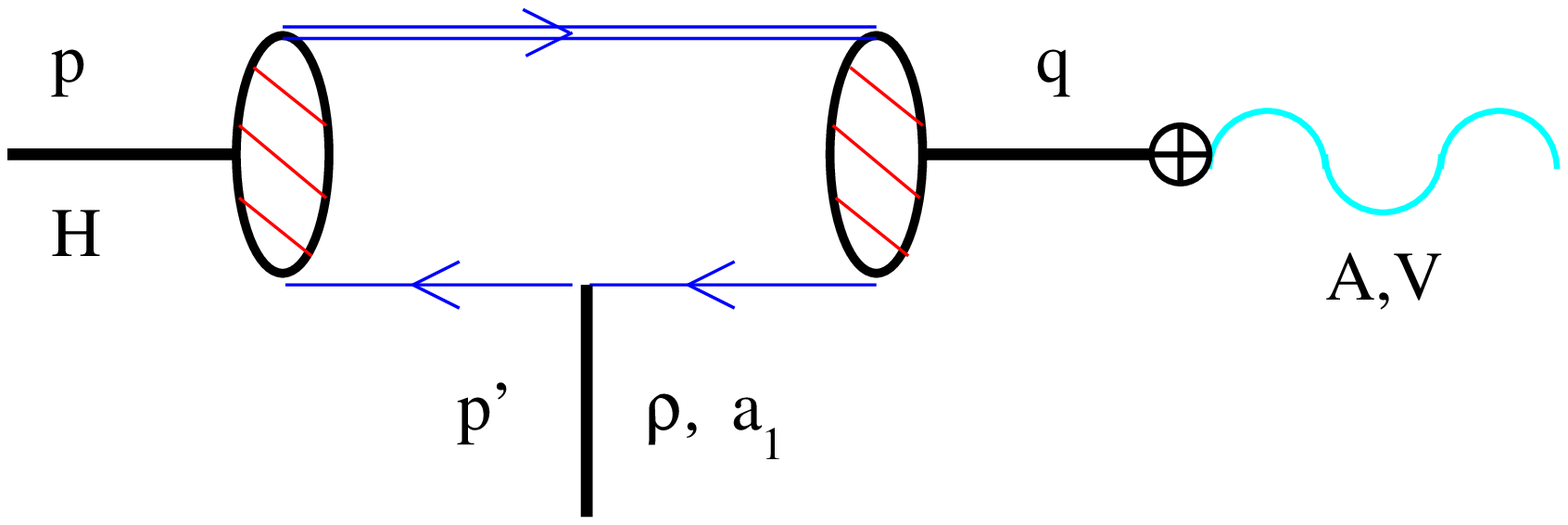}}
\noindent
{\bf Fig. 2} - {Diagram for the polar contribution to the form factor
$B \to \rho$ and $B \to a_1$.}
\end{figure}

\vskip2truecm
\begin{figure}
\epsfxsize=13truecm
\centerline{\epsffile{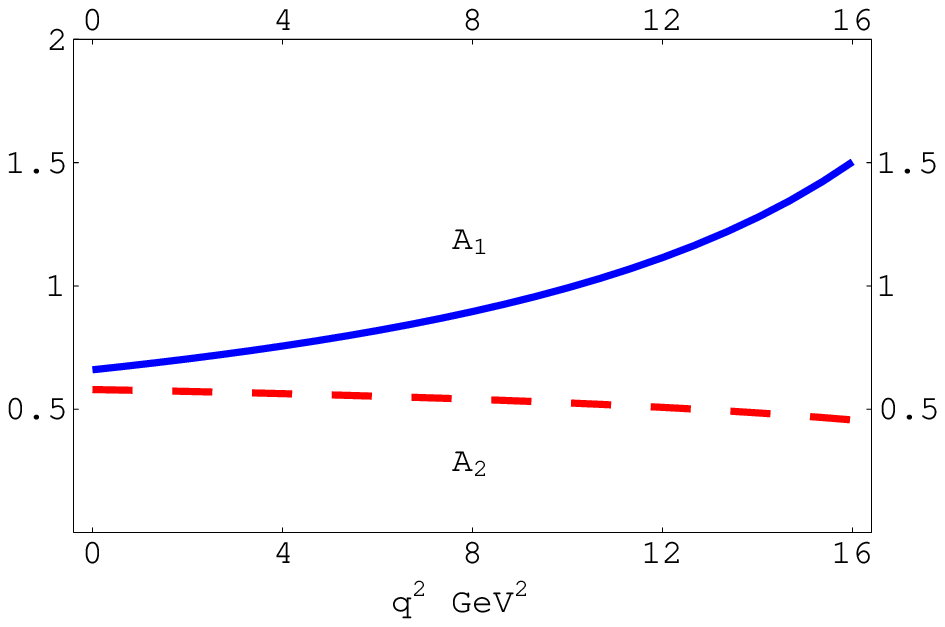}}
\noindent
{\bf Fig. 3} - {Form factors $A_1$ (continuous line) and $A_2$ (dashed line) 
for the semileptonic decay $B \to \rho \ell \nu$.}
\end{figure}
\vskip2truecm

\begin{figure}
\epsfxsize=13truecm
\centerline{\epsffile{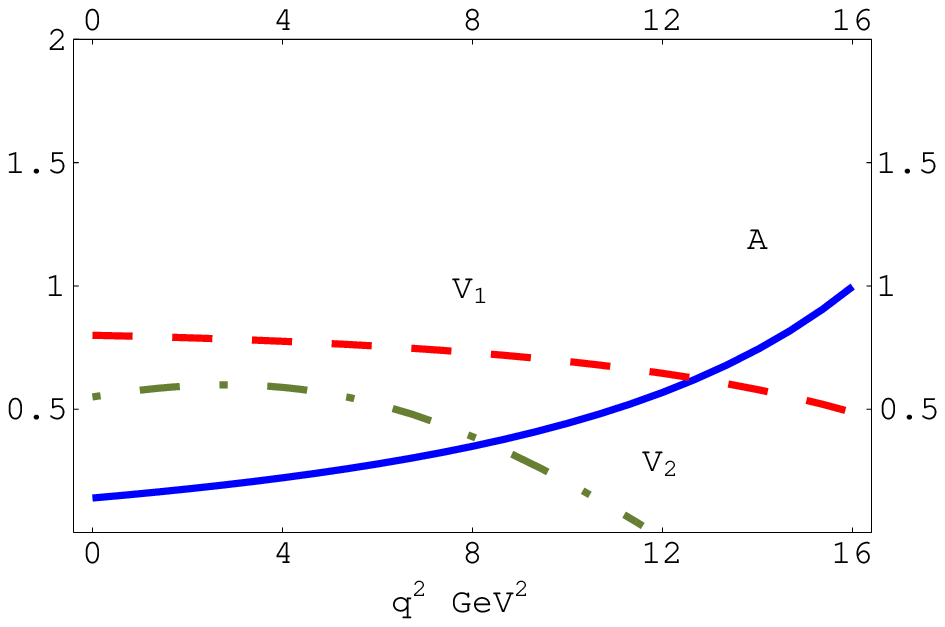}}
\noindent
{\bf Fig. 4} - {Form factors $A$ (continuous line), $V_1$ (dashed line) 
and $V_2$ (dot-dashed line) for the semileptonic decay $B \to a_1 \ell \nu$.}
\end{figure}

\newpage
\begin{table}
\begin{center}
\vbox{\offinterlineskip
\halign{&#& \strut\quad#\hfil\quad\cr
\tableline
\tableline
&  && $V^D$ && $A^D_1$ && $A^D_2$ && $A^D_0$ && $A^D$ && $V^D_1$ && $V^D_2$&&
$V^D_0$ &\cr
\tableline
\tableline
& $F^D(0)$ && $0.83$ && $0.69$ && $0.81$ && $0.33$ && $1.62$ && 
$1.13$ && $1.13$ && $1.13$ &\cr 
& $a_F$ && $0.93$ && $0$ && $0.87$ && $2.9$ && $1.13$ && $0.18$ && 
$1.3$ && $1.9$ &\cr
& $b_F$ && $0.02$ && $0$ && $-0.17$ && $2.6$ && $0.12$ && $0.04$ && 
$3.8$ && $0.93$ &\cr
\tableline
\tableline}}
\caption{Parameters of the direct contribution to the various 
$B$ form factors for $B\to \rho$ and $B\to a_1$ decays. The values $F^D(0)$
for the $B \to a_1$ transition (last four columns) should be multiplied by 
the factor $0.25~{\mathrm GeV}^2 /f_a$. The theoretical 
uncertainty is $\pm 15 \%$.}
\label{t:tab1}
\end{center}
\end{table}

\begin{table}
\hfil
\vbox{\offinterlineskip
\halign{&#& \strut\quad#\hfil\quad\cr
\tableline
\tableline
& && $V^P$ && $A^P_1$ && $A^P_2$ && $A^P_0$ && $A^P$ && $V^P_1$ && $V^P_2$ && 
$V^P_0$ &\cr
\tableline
\tableline
& $F^P(0)$ && $-0.84$ && $-0.11$ && $-0.15$ && $-0.019$ && $-1.48$ &&
 $-0.32$ && 
$-0.57$ && $0.07$ &\cr 
& $m_P$ && $5.3$ && $5.5$ && $5.5$ && $-$ && $5.5$ && $5.3$ && 
$5.3$ && $-$ &\cr
\tableline
\tableline}}
\caption{Parameters of the polar contribution to the various $B$ form 
factors for $B\to \rho$ and $B\to a_1$ decays. Pole masses in GeV.
The values $F^P(0)$
for the $B \to a_1$ transition (last four columns) should be multiplied by 
the factor $(0.25~{\mathrm GeV}^2/f_a)$. The theoretical 
uncertainty is $\pm 15 \%$.}
\label{t:tabp}
\end{table}

\begin{table}
\hfil
\vbox{\offinterlineskip
\halign{&#& \strut\quad#\hfil\quad\cr
\tableline
\tableline
&$\Delta_H$&& ${\hat F}$&& ${\hat F}^+$&\cr
\tableline
&$0.3$&& $0.33$&& $0.22$&\cr
&$0.4$&& $0.34$&& $0.24$&\cr
&$0.5$&& $0.37$&& $0.27$&\cr
\tableline
\tableline}}
\caption{${\hat F}$ and ${\hat F}^+$ for
various values of $\Delta_H $. $\Delta_H$ in GeV, leptonic constants
in GeV$^{3/2}$.}
\label{t:tabf}
\end{table}

\begin{table}
\hfil
\vbox{\offinterlineskip
\halign{&#& \strut\quad#\hfil\quad\cr
\tableline
\tableline
& && This work && Potential Model \cite{ladisa} && LCSR \cite{ballbraun} && 
SR \cite{ballcol} && Latt. + LCSR \cite{lattice} &\cr
\tableline
\tableline
& $V^{\rho}(0)$ && $-0.01 \pm 0.25$ && $0.45~\pm~0.11$ && $0.34~\pm~0.05$ && 
$0.6~\pm~0.2$ && $0.35^{+0.06}_{-0.05}$ &\cr
& $A_1^{\rho}(0)$ && $0.58\pm 0.10$ && $0.27~\pm~0.06$ && $0.26~\pm~0.04$ &&
$0.5~\pm~0.1$ && $0.27^{+0.05}_{-0.04}$ &\cr
& $A_2^{\rho}(0)$ && $0.66\pm 0.12$ && $ 0.26~\pm~0.05$ && 
$0.22~\pm~0.03$ && $0.4~\pm~0.2$ && $0.26^{+0.05}_{-0.03}$ &\cr
& $A_0^{\rho}(0)$ && $0.33 \pm 0.05$ && $0.29~\pm~0.09$ &&  && $ 0.24\pm 0.02$ 
&& $0.30^{+0.06}_{-0.04}$ &\cr
\tableline
\tableline}}
\caption{Form factors for the transition $B \to \rho$ at $q^2=0$. Our results
are compared with the outcome of other theoretical calculations:
potential models, light cone sum rules (LCSR), QCD sum rules (SR), calculations
involving both lattice and light cone sum rules. The large error of $V^\rho (0)$
in our approach is due to the large cancellation between the direct and polar 
contribution.}
\label{t:tab2}
\end{table}

\end{document}